# Single-Sided Contacting of Out-of-Plane Polarized Piezo Films for Fluid Membrane Lenses

M. Stürmer, M. C. Wapler, U. Wallrabe
University of Freiburg, Department of Microsystems Engineering – IMTEK,
Laboratory for Microactuators, Freiburg, Germany

**Abstract:**

We present an adaptive lens which tunes its focal length with a glass membrane that is deformed by an integrated piezo bending actuator. A particular challenge in the fabrication of this type of actuator is the need to contact the electrodes on both sides of the piezo element, i.e. also on the glass surface. Therefore, we show a novel contacting method where the top electrode on the piezo is segmented and the backside electrode is left at a floating potential. With this setup we achieve tuning capabilities of more than 13 m$^{-1}$ and response times in the range of less than 4 µs at a usable aperture of 4.5 mm.

**Motivation and concept**

Tunable fluid-membrane lenses (e.g. [1-4]) commonly use elastomer membranes like silicone as an optical surface and are filled with an optical fluid [3]. Glass membranes, however, promise a wider range of compatible filling liquids due to much greater chemical stability. Therefore, high refractive index liquids can be used, and matching to the glass promises reduced distortions and aberrations. In contrast to other varifocal lenses with glass membranes [5], our approach uses a bulk piezo material, such that larger actuators and thicker, larger membranes can be realized. In the system design in Fig. 1, the piezo actuators are directly attached to the glass which leads to a common problem of bending actuators: The passive glass layer obstructs access to the backside-electrode of the piezo with a wire. A well-known method for contacting piezo films from one side is driving an in-plane polarized piezo with interdigitated surface electrodes in the longitudinal mode [6, 7]. The actual three-dimensional field distribution below the electrodes is, however, far from homogeneous, which leads to a drop in efficiency [7].

As an alternative, we demonstrate here a method to contact a transversely polarized piezo bending actuator with planar electrodes from one side only which leads to an almost perfectly homogeneous field distribution. We separate the electrodes on the outer side in two rings and leave the back electrode at a floating potential that is obtained from the voltage between the front-electrodes and the electrode areas by considering a series connection of capacitors:

$V_{\text{float},1} = V_1 \cdot A_1/(A_1 + A_2)$ (cf. quantities in Fig. 1 bottom). The resulting field is used both to re-polarize the material in the depicted directions and for the actuation.

**Design**

The working principle of the lens design uses the constant volume of the fluid which is encapsulated by the lens membrane. The ring-shaped actuator bends the outer part of the membrane to displace a fluid volume, and the resulting pressure causes it to bulge in the centre, shaping a lens profile.

We chose equal electrode areas, $A_1 = A_2$ to obtain a homogenous strain and maximum overall dielectric displacement over the entire piezo layer. With a simplified linear FEM simulation model we found values for the parameters shown in Fig. 2 top that maximize the achievable deflection of the membrane. We further found that a sandwich design with actuators on both sides of the glass provides a higher displacement than actuator sheets on only one side. A gap between outer support ring and piezo ($r_{\text{ring}} - r_{\text{outer}}$) of 500 µm provides sufficient flexibility to act as a hinge while being strong enough to resist vertical displacement from the counter pressure. Varying the inner and outer radius ($r_{\text{aperture}}$ and $r_{\text{outer}}$) (Fig. 2 bottom) within their manufacturable range, we found that the maximum refractive power is achieved at an outer radius of 11 mm and aperture width of 9 mm when the ring radius is kept constant at 11.5 mm. We used these values to manufacture the prototype.

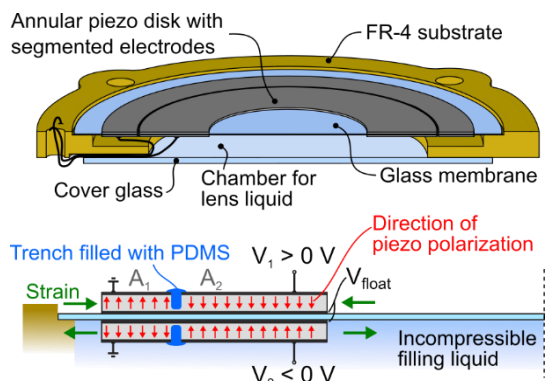

***Fig. 1: Top:*** *Schematic cross-section drawing of the lens.* ***Bottom:*** *Working principle.*

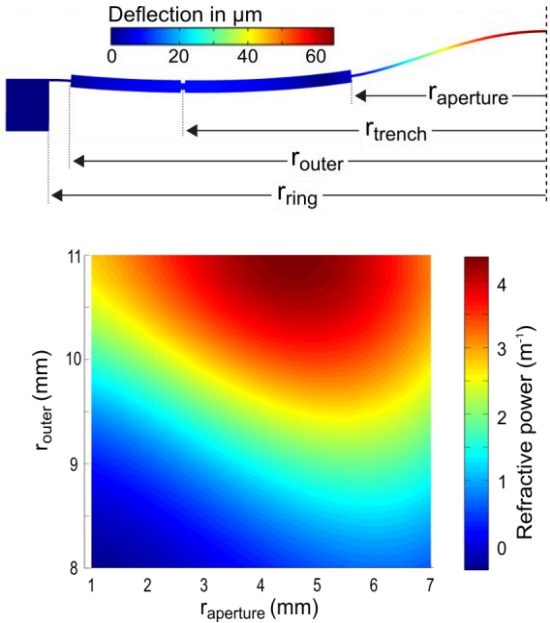

*Fig. 2 Top: Simulated deflection profile (exaggerated by factor 15) with geometry parameters.*
*Bottom: FEM Simulation results of a parameter sweep of aperture width and outer piezo radius at $r_{ring}$ = 11.5 mm and $V_{total}$ = 266 V for a double-sided piezo configuration.*

### Fabrication

The piezo is a transversely polarized 100 μm thick commercial PZT film with planar electrodes and a piezoelectric coefficient $d_{31}$ ~ -270 pm/V. After laser-cutting the piezo layers to an annular shape, they are depolarized at 430°C on a hot plate and glued to the 50 μm thick glass membranes. Afterwards, we laser-structure a 100 μm wide and approx. 60 μm deep insulation trench into the PZT surface (Fig. 1) and cut the glass to a round shape of 25 mm. Remaining debris from the laser processing is removed with a dip in $HNO_3$ in weak ultrasound. We then fill the trench with silicone to avoid electrical breakdown (Momentive RTV 615 with 20 kV/mm dielectric strength), and glue the piezo-glass-membrane stacks to a frame from FR-4 PCB material. The piezo segments are connected with thin wires to the PCB frame by manual soldering. We then re-polarize the piezo films at $E$ ~ 1.5 kV/mm at a temperature above the glass-transition of the glue to reduce the pre-deflection due to the remanent strain. Finally, the lens is filled with paraffin oil with a refractive index of n = 1.48 at 520 nm and 25 °C and is sealed with a 300 μm cover glass on its back side. We built three prototypes: one with piezo layers on both sides of the membrane (which is depicted in Fig. 3) and two single-sided versions with the piezo mounted above or below the membrane.

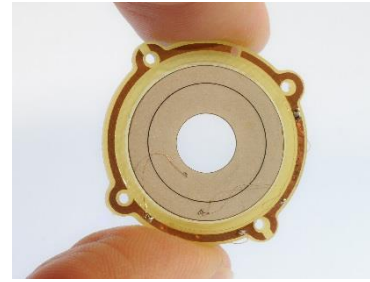

*Fig. 3: Photograph of a prototype.*

### Characterization

The lens prototypes were characterized by scanning the surface profile of the glass membrane with an optical profilometer using a chromatic confocal sensor. The lenses were driven in quasi-static regime (at 0.75 Hz triangular ramp) and with maximum fields ranging from – 0.3 kV/mm (~ 1/3 $E_c$) to 1.5 kV/mm which we consider to be near saturation.

First, we investigated the apertures of the lenses, i.e. the diameter which is usable as a lens. Therefore, we measured the lens surface profile along a line through the membrane centre and fitted a second order polynomial function to the data for varying diameters. The residuals of the regression that are depicted in Fig. 4 are a measure of how well the data can be approximated by a parabola. The plot shows a steep increase of the residuals for apertures larger 5 to 7 mm, as it is also expected from the simulated profile in Fig. 2 because of the change in curvature towards the outer part of the membrane (around the transition from yellow to green). The aperture diameter in an imaging system should therefore be limited to approximately half the membrane diameter to obtain good parabolic lens properties. The residuals for small apertures give an approximation for the limits induced by noise of the measurement method.

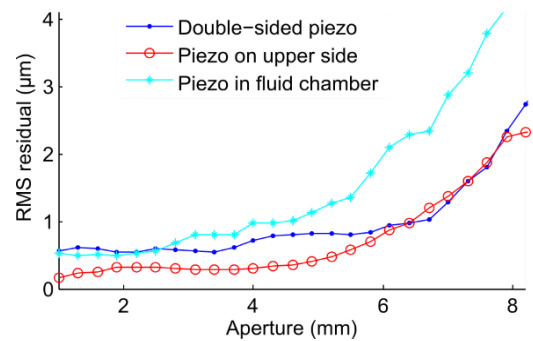

*Fig. 4: Root mean square of the residuals of a quadratic regression of the membrane profile with varying aperture. All lenses operated with the maximum field strength.*

We obtained the refractive power characteristics of the three different lens prototypes by using half the membrane diameter for a spherical regression and applying the Lensmaker's Formula.

We find in Fig. 5 (top) that the double sided layout yields the highest refractive power tuning capability of $f^{-1} > 13$ m$^{-1}$ as expected from simulations, because of the higher stiffness and force that the two actuators provide.

While the two single sided layouts feature a smaller tuning range of 3-4 m$^{-1}$, they are easier to fabricate.

The sign of the displacement depends on whether the piezo is mounted on top or below the membrane. The different shape of the curves compared to the double-sided layout may also come from the different stiffness of the actuators, in particular compared to the hinge area outside of the piezo rings.

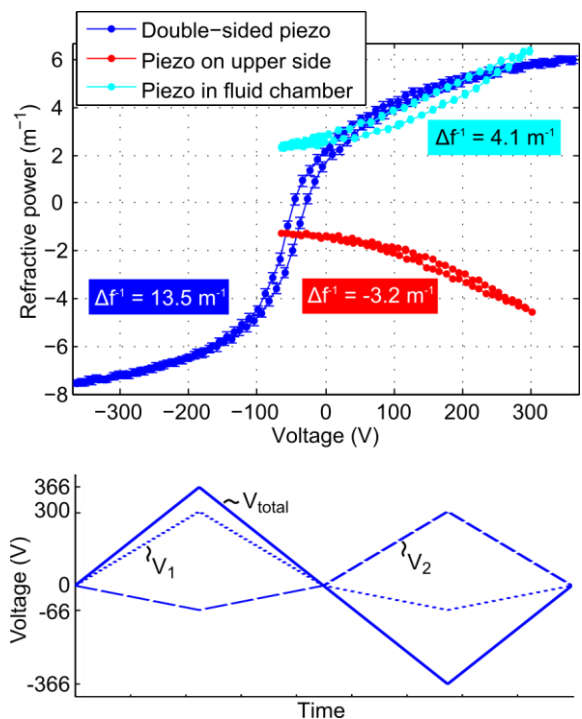

*Fig. 5 Top: Refractive power as a function of the applied voltage for different actuator configurations. Errorbars indicate the regression parameter uncertainty.*
*Bottom: Voltage profile for the operation of the double-sided device. $V_{total} = V_1 - V_2$ is used for plotting in top figure.*

There is a small difference in the behaviour of the lens with two piezo layers for negative and positive actuation voltages. Negative voltages indicate that the bottom actuator inside the fluid chamber is driven in polarization direction while positive voltages mean the same for the upper actuator as depicted in Fig. 5 bottom. Therefore, the shape of the curve for the lens with double-sided actuator is caused by a different contribution of the two actuators to the total deflection. This behaviour is reproducible and can be attributed to non-linear effects caused by counter-pressure in the fluid chamber and to a pre-deflection of the membrane induced by residual stress after the polarization procedure.

One key advantage of adaptive lenses, especially when driven by piezo actuators, is their fast response time compared to conventional lens systems. We therefore characterized the speed of the lens with double-sided actuator as shown in Fig. 6. First, we measured the frequency response of the deflection of the membrane centre point which correlates over a wide range approximately linearly to the refractive power. The system was excited with a sinusoidal signal of different amplitude on the bottom actuator. On the one hand, we see a low pass behaviour that is caused by the damping of the filling liquid (Fig. 6 top). This can be avoided by choosing an oil with lower viscosity. The actual limitation is the first system resonance that is seen at around 1.2 kHz, mostly independent of the excitation amplitude. It represents the fundamental vibration mode, i.e. the main spherical lens deflection and could thus be used for resonant tuning of the focal point.

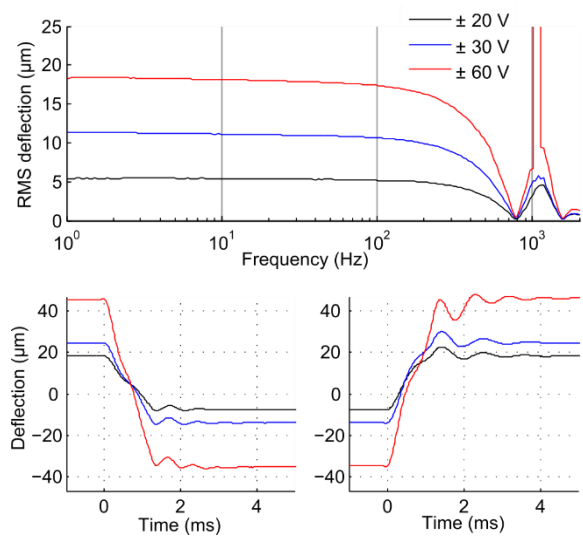

*Fig. 6 Top: Frequency spectrum upon excitation.*
*Bottom: Switching with a step function at t = 0 ms.*

This agrees very well with the step response (upwards and downwards in Fig. 6 bottom), where we see a damped response, with a ringing at approx. 1 kHz. Driving with a step function is highly relevant for example if the lens is used in an autofocus system where the focal length is shifted in discrete steps. While the system settles after around 4 ms, a smart driving circuitry could push the response time to around 1 ms.

The optical transmittance was measured with a photo-spectrometer and averaged over three prototypes. The transmission is above 90% throughout the visible wavelength range as shown in Fig. 7.

It is mostly limited by the Fresnel reflections at the air – glass interface which could, due to the glass surface, still be improved with an anti-reflection coating on the membrane and the cover plate on the back side. The reflections between glass and oil, however, are already low because of similar refractive indices of those materials. They could be even further reduced using a fully index-matched filling liquid.

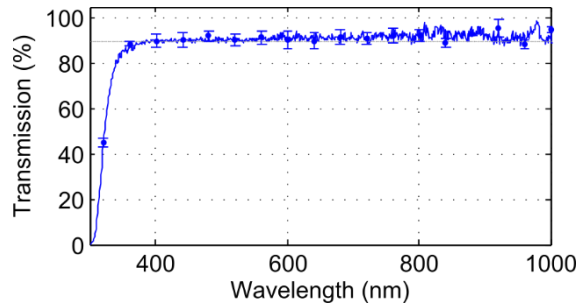

*Fig. 7: Averaged transmission spectrum measured for three lenses filled with paraffin oil with a photospectrometer.*

**Conclusions**

We presented an adaptive lens that uses a deformable glass membrane. To realize the integrated actuation, we glued a bulk piezo actuator onto the membrane which requires a new method to contact the electrodes, as the inner electrodes are obstructed by the glass membrane. Rather than using a complicated process to contact the backside-electrodes via the glass membrane, we have successfully demonstrated a method to contact transversely polarized piezo films from one side only. We derived a lens design from FEM simulations and built prototypes. The use of a glass membrane allows for a wide range of filling liquids, reasonably large focal power tuning capability of more than 13 m$^{-1}$ at usable apertures of more than 4 mm and very fast response times in the range of some milliseconds.

Future work will include a robust fabrication process as well as optical aberration correction at higher orders than the aspherical correction that we demonstrated with a slightly different working principle in [8,9].

**Acknowledgements**

This research was supported by German Research Foundation (DFG) grant WA 16547/1-2 within the Priority Program "Active Micro-optics" and the Cluster of Excellence BrainLinks-BrainTools EXC 1086.

**Contact**

U. Wallrabe; wallrabe@imtek.uni-freiburg.de